\documentclass[epj]{svjour}
\usepackage{graphicx}
\begin{document}
\title{Understanding nucleon structure using lattice simulations}
\subtitle{Recent progress on three different structural observables}
\author{Wolfram Schroers}
\institute{John von Neumann-Institut f\"ur Computing NIC/DESY, 15738 Zeuthen, Germany}
\date{Received: date / Revised version: date}
%
\abstract{This review focuses on the discussion of three key results
  of nucleon structure calculations on the lattice. These three
  results are the quark contribution to the nucleon spin, $J_q$, the
  nucleon-$\Delta$ transition form factors, and the nucleon axial
  coupling, $g_A$. The importance for phenomenology and experiment is
  discussed and the requirements for future simulations are pointed
  out. \\ Preprint number: DESY 06-194 \\ \\
} 
\maketitle
\section{Introduction\label{sec:introduction}}
In recent years lattice gauge theory has become a mature and reliable
way to investigate the structure of strong interactions. It provides a
model-independent way to do calculations in QCD\@. However,
contemporary lattice computations become extremely costly at quark
masses corresponding to pion masses below $500$ MeV. Nature, however,
has chosen the pion mass to be only $140$ MeV. The lightness of the
pseudoscalar mesons is due to the mechanism of spontaneous chiral
symmetry breaking. If, however, we can investigate only the regime of
heavy quarks, where chiral symmetry is broken explicitly by the quark
mass, we might not describe physics accurately at light quark masses.

To address and overcome this challenge, three different procedures
have been proposed and are actively pursued: (i) Pushing existing
simulations with Wilson-type quarks down to smaller quark masses by
relying on improved algorithms and faster
computers~\cite{Jansen:2006pr}, (ii) using a hybrid action approach by
using different formulations for sea- and
valence-quarks~\cite{Negele:2004iu}, and (iii) doing simulations using
dynamical Ginsparg-Wilson formulations, such as Domain-Wall
fermions~\cite{Tweedie:2006la} or Overlap
fermions~\cite{Cundy:2004xf}. The last approach is certainly the most
challenging and demanding one since the entire parameter space has to
be explored again. This applies also to heavy quarks, a regime in
which Ginsparg-Wilson fermions are about $30$ to $100$ times more
expensive than standard Wilson-type fermions.

The hybrid action ansatz is an excellent compromise between quark mass
and performance, but suffers from conceptual problems. First of all,
the hybrid theory breaks unitarity at finite lattice spacing. Thus, it
cannot act as an effective theory at finite lattice spacing, and the
existence of the continuum limit is crucial. Furthermore, usually
staggered-type quarks are being used for the sea with the square-root
being taken of the determinant. It is not clear if the procedure of
taking the square root commutes with taking the continuum limit, see
e.g.~\cite{Durr:2005ax} for a recent review. Finally, the matching of
sea- and valence-quark masses is prescription-dependent, and
particular choices may give rise to additional possibly large ${\cal
  O}(a^2)$ artifacts~\cite{Bar:2005tu}.

In this review we focus on three observables with relevance to
phenomenological and experimental applications. The first one is the
quark contribution to the nucleon spin, $J_q$. The second one is the
transition form factors of the nucleon-$\Delta$ transition.  The third
one is the nucleon axial coupling, $g_A$. The former two of these
quantities have so far been understood qualitatively, but a precise
matching between the light quark regime and the lattice --- possibly
by chiral perturbation theory or an effective model of the strong
interaction like~\cite{Goeke:2005fs} --- still remains to be done. For
the latter observable it has been shown that lattice data can in fact
be consistent with experiment when fitting it using the leading
logarithmic chiral perturbation theory expression. This achievement
marks a milestone in the field of nucleon structure.

\section{Quark contribution to nucleon spin\label{sec:quark-contr-nucl}}
The quark contribution to the spin of the nucleon has been under
intense scrutiny after the observation that only about $(20\pm 15)\%$
of the nucleon spin arises from the quark spin~\cite{Ball:1995td}.
Recently, it has been realized how the use of
GPDs~\cite{Mueller:1998fv,Radyushkin:1997ki,Ji:1996ek} provides the
means to directly compute the quark contribution to the nucleon spin
via the energy momentum tensor~\cite{Ji:1996ek}
\begin{equation}
  \label{eq:jqdef}
  J_q = \lim_{t\to 0} \left(A^{\mbox{\tiny u+d}}_{20}(t) +
  B^{\mbox{\tiny u+d}}_{20}(t)\right)\,.
\end{equation}
The virtuality $t$ is given by $t\equiv( p'-p)^2$, where $p'$ and $p$
are the nucleon's incoming and outgoing momenta. The generalized form
factors, $A^{\mbox{\tiny u+d}}_{20}(t)$ and $B^{\mbox{\tiny
    u+d}}_{20}(t)$, show up in the parameterization of the nucleon's
energy-momentum tensor. For further details and the exact definition
consult~\cite{Ji:1996ek}. The challenge is to understand which
fraction of the nucleon spin, $J_N=1/2$, arises from the quark spin,
$1/2\Sigma_q$, the quark orbital angular momentum, $L_q$, and which
fraction comes from gluon contributions, $J_g$:
\begin{equation}
  \label{eq:jqpart}
  J_N = 1/2 = J_q + J_g = 1/2\Sigma_q + L_q + J_g\,.
\end{equation}
The value of $\Sigma_q$ has been known before~\cite{Mathur:1999uf}.
The new ingredient is the ability to directly calculate $J_q$, and
thus also $L_q$. To this end, there is no experimental determination
of that quantity. The first computation of $J_q$ on the lattice has
been done in \cite{Mathur:1999uf}. This calculation only utilizes
quenched Wilson fermions, but features a calculation of the
disconnected contribution using noisy estimators. A later calculation
\cite{Gockeler:2003jf} calculates all generalized form factors in the
energy-momentum tensor separately and at the same time a publication
\cite{Hagler:2003jd} features full QCD and introduces an improved
technology to extract form factors from matrix elements. Higher
moments of GPDs have also been computed~\cite{Hagler:2003is}.

As of today, the understanding gained from the world of pions weighing
$500$ MeV and beyond is that the quark contribution to the nucleon
spin is about $70\%$, all of which comes from the quark spin alone.
The remaining $30\%$ comes from the gluons. The quark orbital angular
momentum is negligibly small due to a cancellation between the
contributions of u- and d-quarks~\cite{Negele:2004iu}.

This result differs from the finding outlined above which indicates
that this quantity can be expected to substantially depend on the pion
mass. The cancellation of the orbital angular momentum for u- and
d-quarks is an interesting qualitative feature. The insight that the
nucleon in the heavy pion world receives a larger fraction of its spin
from quarks rather than gluons is compatible with expectations from
the non-relativistic quark model, but the exact interpolation between
the heavy quark and the light quark regime can give further insight
into how the strong interaction operates.

The extrapolation to the chiral regime, however, has not yet been
possible and hence a precise quantitative matching with Nature has not
yet been established. Although Ref.~\cite{Chen:2001pv} suggests a
rather flat expression it is yet unclear whether the same straight
line is to be used for the light quark regime as the one fitting the
simulations. In this situation it is inevitable to perform similar
calculations at smaller pion masses before a matching between lattice
and small-scale expansion schemes can be established and a definitive
prediction from the lattice can be provided.

Further investigations from several groups are underway and all three
different paths outlined in Sec.~\ref{sec:introduction} are taken to
resolve this important question. We can conclude, however, that the
technology and understanding of how to compute these matrix elements
are available and can be deployed easily once sufficiently light pion
masses are available.

\section{$N\rightarrow\Delta$ transition form
  factors\label{sec:nright-trans-form}} A key question is whether the
baryon states of QCD are spherical or deformed. Although the nucleon
is easily accessible in exclusive and inclusive scattering
experiments, it cannot have a spectroscopic quadrupole moment since it
has spin $J_N=1/2$. The excited states with spin 3/2 and above can
have a quadrupole moment, but these are not easily accessible in
experiments.  The only way to learn about deformations of the
low-lying baryon spectrum is to consider transitions between the
nucleon and the first excited state, the $\Delta(1232)$ resonance.
Experimentally, a flurry of activity has recently lead to several
important and exciting results~\cite{Blanpied:1996yh}.

The nucleon-$\Delta$ transition can be parameterized using three form
factors --- the dominant magnetic dipole form factor, ${\cal G}_{M1}$,
the electric quadrupole, ${\cal G}_{E2}$, and the Coulomb quadrupole,
${\cal G}_{C2}$. Should the nucleon-$\Delta$ system be deformed, the
latter two form factors will not vanish. Should the system be
spherical, only the magnetic dipole form factor will be non-zero.

On the lattice, publications reporting the successful computation of
these transition form factors are~\cite{Alexandrou:2003ea}. This set
of calculations used unquenched Wilson fermions with pion masses
beyond $600$ MeV and quenched Wilson ferm\-i\-ons with pion masses larger
than $370$ MeV. Later it has been attempted to apply these techniques
also for hybrid actions~\cite{Alexandrou:2005em}, but to this end the
statistical error bars on the quadrupole form factors turn out to be
too large.

From these studies it has been clearly established, that the
nucleon-$\Delta$ system is indeed deformed. The sign and the order of
magnitude of the quadrupole form factors ${\cal G}_{E2}$ and ${\cal
  G}_{C2}$ was extracted successfully. The heavy pion world in fact is
similar to Nature for these observables.

However, the extrapolation to the physical pion masses yields an
inconsistency for ${\cal G}_{C2}$ at values below $Q^2<0.2\,$GeV$^2$.
This discrepancy has been addressed recently in the framework of
chiral perturbation theory~\cite{Pascalutsa:2005ts}. It appears
plausible that the discrepancy in fact arises from the inadequacy of a
linear chiral extrapolation --- it still remains to be seen if lattice
data at smaller pion masses can indeed verify the pion mass dependence
suggested in~\cite{Pascalutsa:2005ts}.

\section{Nucleon axial coupling $g_A$\label{sec:nucl-axial-coupl}}
The investigation of the nucleon axial coupling has a long history on
the lattice, see~\cite{Schroers:2005rm} for recent reviews. Several
groups have performed investigations using a wide array of different
lattice actions, spacings, volumes, and pion masses.

\begin{figure}
  \begin{center}
    \includegraphics[scale=0.3,angle=270]{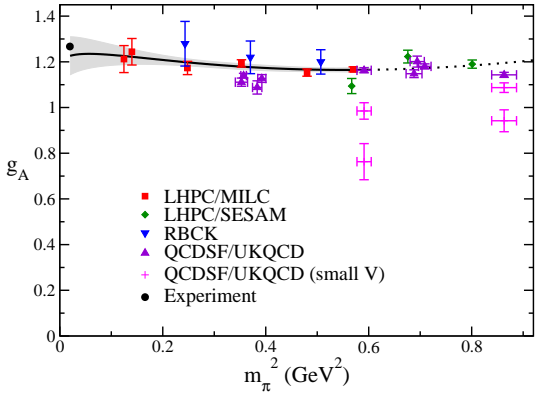}
    \vspace{-0.1in}
  \end{center}
  \label{fig:lhpcga}
  \caption{Full QCD computations of the nucleon axial coupling,
    $g_A$. The line shows the fit of the leading logarithmic $\chi$PT
    expression to the hybrid lattice data from the LHPC collaboration.
    Results from other groups are plotted, but not included in the
    fit. Figure taken from Ref.~\cite{Edwards:2005ym}, QCDSF data
    updated from Ref.~\cite{Khan:2006de}.}
\end{figure}

Recently, two independent papers~\cite{Edwards:2005ym}
and~\cite{Khan:2006de} have appeared showing how current lattice data
can in fact be combined with chiral perturbation theory to arrive at
the experimental value.  Figure~\ref{fig:lhpcga} shows the application
of the leading logarithmic expression from $\chi$PT to the hybrid data
computed by the LHPC collaboration in~\cite{Edwards:2005ym}. The
gray-shaded error band shows the error arising from statistical
uncertainties only. The fit yields quantitative agreement with the
experimental value.

However, the applicability of the leading order chiral perturbation
theory expression to the pion masses available has been questioned
in~\cite{Bernard:2006te}. The flat behavior at pion masses beyond
$300$ MeV is attributed to fine-tuning between different terms in the
expansion.
On the other hand, the expansion can still be consistent with
experiment when applied to lattice calculations employing pion masses
as large as $600$ MeV~\cite{Khan:2006de}. It is perhaps fair to say
that the exact range of applicability of $\chi$PT is under debate.
Nonetheless, the striking agreement between the fit of lattice data
and the experimental value mark an important milestone for the lattice
treatment of nucleon structure.

\section{Summary\label{sec:summary}}
We have given three examples of recent lattice calculations which are
of great interest to both phenomenologists and experimentalists alike.
The limiting factor of all these lattice results, however, is their
limitation to rather large quark masses. Currently, the question of
chiral extrapolations is under debate and the applicability depends
strongly on the observable. While some groups successfully apply fits
to pion masses as large as $600$ MeV, other groups believe that pion
masses lower than $300$ MeV are essential. While the latter mass
regime has not been reachable so far, we believe that the upcoming
generation of lattice calculations will be able to settle the debate.

This work was supported in part by the DFG, contract FOR 465 (FG
Gitter-Hadronen-Ph{\"a}nomenologie), and in part by the EU Integrated
Infrastructure Initiative Hadron Physics (I3HP), contract number
RII3-CT-2004-506078.


\begin{thebibliography}{99}
\bibitem{Jansen:2006pr} K.~Jansen {\it et al.}, these proceedings;
  M.~G{\"o}ckeler {\it et al.}, arXiv:hep-lat/0610066.
\bibitem{Negele:2004iu} J.~W.~Negele {\it et al.}, Nucl.\ Phys.\ 
  Proc.\ Suppl.\ {\bf 128} (2004) 170;
  D.~B.~Renner {\it et al.}  [LHP Collaboration], Nucl.\ Phys.\ Proc.\ 
  Suppl.\ {\bf 140} (2005) 255;
  Ph.~H{\"a}gler, J.~W.~Negele, D.~B.~Renner, W.~Schroers, T.~Lippert
  and K.~Schilling [LHPC Collaboration], Eur.\ Phys.\ J.\ A {\bf 24S1}
  (2005) 29;
  R.~G.~Edwards {\it et al.}  [LHPC Collaboration], PoS {\bf LAT2005}
  (2006) 056;
  R.~G.~Edwards {\it et al.}, arXiv:hep-lat/0610007.
\bibitem{Tweedie:2006la} R.~Tweedie {\it et al.}  [UKQCD and RBC
  Collaborations], PoS {\bf LAT2005} (2006) 096.
\bibitem{Cundy:2004xf}
  N.~Cundy, S.~Krieg, A.~Frommer, T.~Lippert and K.~Schilling,
  Nucl.\ Phys.\ Proc.\ Suppl.\  {\bf 140}, 841 (2005);
  N.~Cundy, S.~Krieg and T.~Lippert, PoS {\bf LAT2005}, 107 (2006);
  S.~Sch{\"a}fer, arXiv:hep-lat/0609063.
\bibitem{Durr:2005ax} S.~D{\"u}rr, PoS {\bf LAT2005}, 021 (2006).
\bibitem{Bar:2005tu} O.~B{\"a}r, C.~Bernard, G.~Rupak and N.~Shoresh,
  Phys.\ Rev.\ D {\bf 72}, 054502 (2005).
\bibitem{Goeke:2005fs} K.~Goeke, J.~Ossmann, P.~Schweitzer and
  A.~Silva, Eur.\ Phys.\ J.\ A {\bf 27}, 77 (2006).
\bibitem{Ball:1995td} R.~D.~Ball, S.~Forte and G.~Ridolfi, Phys.\ 
  Lett.\ B {\bf 378} (1996) 255.
\bibitem{Mueller:1998fv}
  D.~M{\"u}ller, D.~Robaschik, B.~Geyer, F.~M.~Dittes and J.~Horejsi,
  Fortsch.\ Phys.\  {\bf 42} (1994) 101.
\bibitem{Ji:1996ek} X.~D.~Ji, Phys.\ Rev.\ Lett.\ {\bf 78} (1997) 610.
\bibitem{Radyushkin:1997ki} A.~V.~Radyushkin, Phys.\ Rev.\ D {\bf 56}
  (1997) 5524.
\bibitem{Mathur:1999uf} N.~Mathur, S.~J.~Dong, K.~F.~Liu,
  L.~Mankiewicz and N.~C.~Mukhopadhyay, Phys.\ Rev.\ D {\bf 62} (2000)
  114504.
\bibitem{Gockeler:2003jf} M.~G{\"o}ckeler, R.~Horsley, D.~Pleiter,
  P.~E.~L.~Rakow, A.~Sch{\"a}fer, G.~Schierholz and W.~Schroers [QCDSF
  Collaboration], Phys.\ Rev.\ Lett.\ {\bf 92} (2004) 042002.
\bibitem{Hagler:2003jd} P.~H{\"a}gler, J.~Negele, D.~B.~Renner,
  W.~Schroers, T.~Lippert and K.~Schilling [LHPC collaboration],
  Phys.\ Rev.\ D {\bf 68}, 034505 (2003).
\bibitem{Hagler:2003is} P.~H{\"a}gler, J.~W.~Negele, D.~B.~Renner,
  W.~Schroers, T.~Lippert and K.~Schilling [LHPC Collaboration],
  Phys.\ Rev.\ Lett.\ {\bf 93} (2004) 112001.
\bibitem{Chen:2001pv} J.~W.~Chen and X.~d.~Ji, Phys.\ Rev.\ Lett.\ 
  {\bf 88} (2002) 052003.
\bibitem{Blanpied:1996yh} G.~Blanpied {\it et al.}  [LEGS
  Collaboration], Phys.\ Rev.\ Lett.\ {\bf 76}, 1023 (1996); R.~Beck
  {\it et al.}, Phys.\ Rev.\ C {\bf 61}, 035204 (2000);
  C.~Mertz {\it et al.}, Phys.\ Rev.\ Lett.\ {\bf 86}, 2963 (2001);
  N.~F.~Sparveris {\it et al.}  [OOPS Collaboration], Phys.\ Rev.\ 
  Lett.\ {\bf 94}, 022003 (2005);
  K.~Joo {\it et al.}  [CLAS Collaboration], Phys.\ Rev.\ Lett.\ {\bf
    88}, 122001 (2002);
  A.~M.~Bernstein, Eur.\ Phys.\ J.\ A {\bf 17}, 349 (2003).
\bibitem{Alexandrou:2003ea} C.~Alexandrou {\it et al.}, Phys.\ Rev.\ D
  {\bf 69}, 114506 (2004);
  C.~Alexandrou, Ph.~de Forcrand, H.~Neff, J.~W.~Negele, W.~Schroers
  and A.~Tsapalis, Phys.\ Rev.\ Lett.\ {\bf 94}, 021601 (2005).
\bibitem{Alexandrou:2005em} C.~Alexandrou {\it et al.}, PoS {\bf
    LAT2005}, 091 (2006).
\bibitem{Pascalutsa:2005ts} V.~Pascalutsa and M.~Vanderhaeghen, Phys.\ 
  Rev.\ Lett.\ {\bf 95} (2005) 232001.
\bibitem{Schroers:2005rm} W.~Schroers, Nucl.\ Phys.\ A {\bf 755}
  (2005) 333;
  W.~Schroers, Nucl.\ Phys.\ Proc.\ Suppl.\ {\bf 153} (2006) 277.
\bibitem{Edwards:2005ym} R.~G.~Edwards {\it et al.}  [LHPC
  Collaboration], Phys.\ Rev.\ Lett.\ {\bf 96} (2006) 052001.
\bibitem{Khan:2006de} A.~Ali Khan {\it et al.}, arXiv:hep-lat/0603028.
\bibitem{Bernard:2006te} V.~Bernard and U.~G.~Meissner, Phys.\ Lett.\ 
  B {\bf 639}, 278 (2006).
\end{thebibliography}
\end{document}